\documentclass[12pt]{elsart} 

\usepackage{amssymb}
\usepackage{graphicx}

\begin{document}


\title{Coherent addition of two dimensional array of fiber lasers}

\author{Moti Fridman}
\author{Vardit Eckhouse}
\author{Nir Davidson}
\author{Asher A. Friesem}
\address{Dept. of Physics of Complex system, Weizmann institute of Science,
Rehovot 76100, Israel}

\author{Elena Luria}
\author{Vladimir Krupkin}
\address{ELOP, ElectroOptics Industries LTD. P.O. Box 1165, Rehovot 76111, Israel}

\begin{frontmatter}


\begin{abstract}

Configurations for efficient free space coherent addition of four
separate fiber lasers arranged in two dimensional array are
presented. They include compact and robust interferometric
combiners that can be inserted either inside or outside the cavity
of the combined lasers system. The results reveal that over 85\%
combining efficiency can be obtained.

\end{abstract}


\end{frontmatter}

\maketitle


\noindent

\section{Introduction}

Several approaches have been developed for phase locking and
coherent addition of fiber lasers in order to generate a combined
laser with high output power and high brightness. Some of these
involve active phase locking~\cite{Fan}, Talbot self
imaging~\cite{Wrage, MultiCoreFibers2}, evanescent wave
coupling~\cite{Michaille}, and spatial
filters~\cite{PhaseLockingChina, PhaseLockingChina2, Loftus}, for
obtaining phase locking within the laser cavity and subsequent
external coherent addition of all phased locked lasers. In these
the brightness of the combined far field spot is relatively low
because of the inherent separations between the lasers. Other
approaches involve direct coherent addition within the lasers
cavities by inserting intra-cavity fiber couplers\cite{Kozlov,
Shirakawa}. Although in these direct approaches the alignment is
relatively easy and the stability is good, the combined output
power is rather limited in order to prevent damage and undesirable
non-linear effects to a single output fiber.

To overcome damage and deleterious non-linearities to the output
fibers, it is possible to resort to configurations in which the
coherent addition is performed in free space with robust
intra-cavity interferometric combiners. Such configurations were
already effectively demonstrated for coherent addition of several
solid state lasers whose cavity lengths and alignments could be
readily controlled~\cite{Liran25, Vardit16}. With fiber lasers,
whose cavity lengths could not be controlled with sufficient
accuracy and alignment is difficult, the situation is much more
complicated. Earlier attempts with configurations in which beam
splitters in free space were exploited, successfully demonstrated
coherent addition of two fiber lasers~\cite{Moti2006,
VarditScience}. Unfortunately, such configurations cannot be
easily upscaled because they are bulky and difficult to align.

In this article, we present two robust, stable, highly efficient,
free-space configurations for coherently adding four fiber lasers
arranged in two dimensional array. In these configurations
coherent addition is performed by means of compact interferometric
combining elements in free space. Such configurations could be
readily upscaled to coherently add many fiber lasers.

%

\section{Experimental Configuration and Procedure}

\begin{figure}[h]
\centerline{\includegraphics[width=8.3cm]{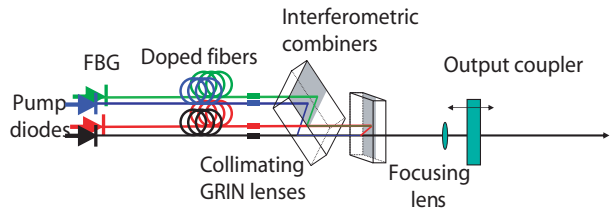}}
\caption{(color online) Basic intra-cavity configuration for
coherent addition of four fiber lasers in free space using two
orthogonally oriented interferometric combiners.} \label {setup}
\end{figure}

We considered two different configurations for free-space coherent
addition of four fiber lasers. The first, namely the intra-cavity
configuration, is presented schematically in Fig.~\ref{setup}. It
included four fiber lasers operating at $1550nm$ and two
intra-cavity interferometric combiners. Each fiber laser consisted
of double-clad single-mode Erbium doped fiber of about 6 meters in
length, where one end was attached to a high reflection fiber
Bragg grating (FBG) of $5nm$ spectral bandwidth around $1550nm$
that served as a back reflector mirror and the other end was
spliced to a collimating graded index (GRIN) lens with
anti-reflection layer to suppress any reflections back into the
fiber cores, and a flat output coupler of 20\% reflection that was
common to all fiber lasers. A focusing lens was inserted in front
of the output coupler for controlling the coupling strength
between the lasers. Each fiber laser was pumped with a multi-mode
diode laser of $915nm$ wavelength, that was spliced to the back of
the FBG. The interferometric combiner was a planar substrate,
where half of the front surface was coated with an antireflection
layer and the other half with a 50\% beam-splitter layer, while
half of the rear surface was coated with a highly reflecting layer
and the other half with an antireflection layer~\cite{Vardit16}.

We measured the spectrum of each fiber laser separately with a
fast photo-detector connected to an RF spectrum analyzer. From
these spectra we determined that the overall lengths of the four
laser cavities were $6.33m$, $6.41m$, $6.52m$ and $6.64m$, and
that the relative intensity noise (RIN) was less than $-70dB/Hz$.
We also measured the output power of each fiber laser as a
function of the diode pump power, and found that the
light-to-light efficiency was about 10\% for each.

\begin{figure}[h]
\centerline{\includegraphics[width=8.3cm]{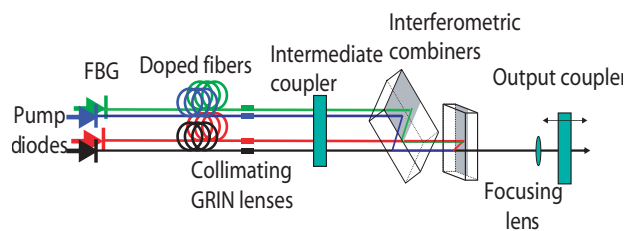}}
\caption{(color online) Basic outer-cavity configuration for
coherent addition of four fiber lasers in free space using two
orthogonally oriented interferometric combiners.} \label {setup2}
\end{figure}

The second configuration, namely the outer-cavity configuration,
is presented schematically in Fig.~\ref{setup2}. It differs from
that shown in Fig.~\ref{setup} in that an intermediate coupler of
20\% was inserted at the output of the fiber lasers, so as to
obtain efficient lasing from each, and the interferometric
combiners were outside the cavities of the four individual lasers.
Coupling between the lasers and consequent phase locking and
coherent addition was achieved by the feedback to each fiber laser
from the common output coupler. In both configurations, when the
fiber lasers are phase locked, the first interferometric combiner
transforms efficiently four light beams to two beams and the
second interferometric combiner transforms the two beams into one
nearly gaussian beam.

To ensure that the beams emerging from the four individual fibers
are exactly parallel, we first aligned and set one fiber in a
mounting arrangement with respect to a planar output coupler so as
to obtain optimal lasing. Next, we inserted the second fiber laser
and aligned it with micrometer stages so it will also optimally
lase from the same output coupler, and bonded it to the same
mounting arrangement with UV cured optical cement. This procedure
was repeated for the third and then fourth fiber lasers. We then
measured the beam parallelism and ascertained that it is better
than $0.1 mrad$. The output light power from the individual fiber
lasers were adjusted to be about the same by controlling their
individual pump powers.

\section{Experimental Results}

\begin{figure*}[h]
\centerline{\includegraphics[width=16.6cm]{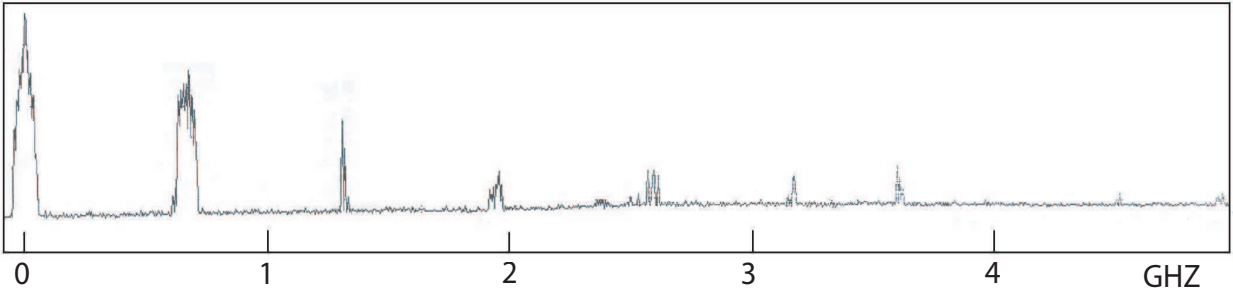}}
\caption{Representative spectrum of the output beam from the
outer-cavity configuration. Clear structure of lobs can be seen,
indicating common longitudinal modes needed for phase locking and
coherent addition.} \label {spectrum}
\end{figure*}

We began our experiments by performing spectral characterization
of the combined output beam using a fast photo-detector connected
to a spectrum analyzer, in order to determine the location and
distribution of the common longitudinal modes~\cite{Moti2006}. A
representative spectrum of the outer-cavity configuration, that
contains four cavities of different lengths, each comprised of
three mirrors, is presented in Fig.~\ref{spectrum}. As evident,
there are many common bands of longitudinal modes that are needed
for phase locking and coherent addition in both configurations,
where the separation between the bands is $700MHZ$ which
corresponds to $20cm$ length difference. This is the distance
between the intermediate coupler and the output coupler. We also
measured the spectrum for the simpler intra-cavity configuration
and verified that there are even more common bands of longitudinal
modes. When upscaling to a large number of fiber lasers, it is
still an open question whether it would be possible to find common
longitudinal modes.

Next we determined the intermediate and final combining
efficiencies in the two configurations, i.e the ratio of the
output power of the combined laser beam over the sum of the output
power of the individual laser beams. This was done by first
inserting only one interferometric combiner in order to coherently
add horizontally the light from the four fiber lasers, each with a
power of $35mW$, to form two beams, each with a power of about
$66mW$. This corresponds to a combining efficiency of over 90\%.
Then, we added the second interferometric combiner to coherently
add vertically the two remaining beams into one beam to obtain a
combined output power of about $121mW$. This corresponds to an
overall combining efficiency (i.e. the ratio between the overall
power with the interferometric combiners to that without the
combiners) of 86\%. The results were essentially the same for both
configurations with a somewhat lower combining efficiency for the
outer-cavity configuration, but still of over 80\%.

\begin{figure}[h]
\centerline{\includegraphics[width=8.3cm]{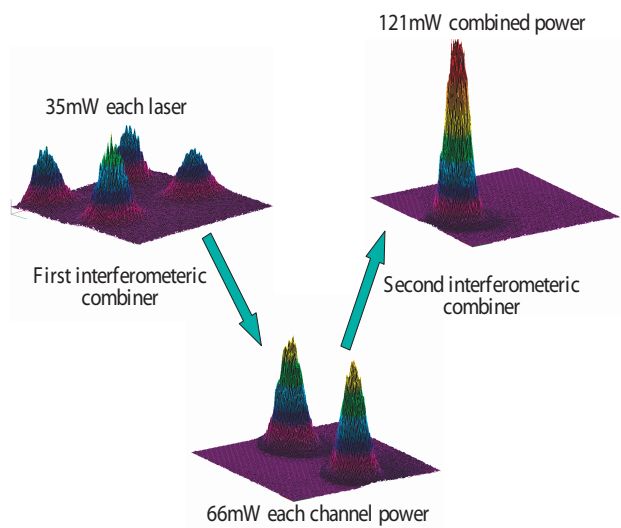}}
\caption{(color online) Experimental intensity distributions for
the intra-cavity configuration. (a) Before the interferometric
combiners; (b) after the first interferometric combiner; (c) after
the second interferometric combiner.} \label {Laser_intensities}
\end{figure}

We also placed MicronViewer 7290 infrared cameras to detect the
intensity distributions before the interferometric combiners,
after the first interferometric combiner and after the second one.
Representative results are shown in Fig.~\ref{Laser_intensities}.
As evident, the initial four beams first coherently transform to
two beams after the first interferometric combiner and then to one
beam after the second interferometric combiner.

\begin{figure}[h]
\centerline{\includegraphics[width=8.3cm]{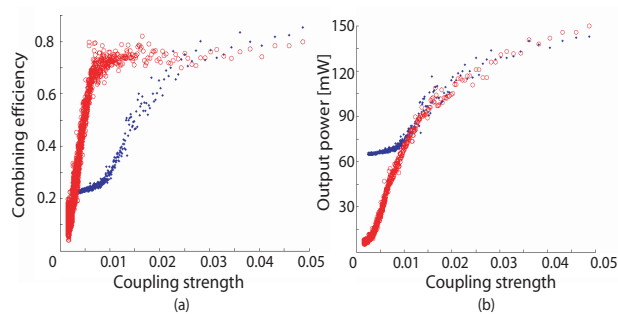}}
\caption{(color online) Experimental combining efficiency and
output power as a function of the coupling strength for the
intra-cavity and outer-cavity configurations. Circles denote the
results for the intra-cavity configuration. Dots denote the
results for the outer-cavity configuration} \label {graphs}
\end{figure}

Finally, we measured the combined output power and determined the
combining efficiency as a function of the coupling strength,
namely the relative amount of light that is re-injected from one
laser into the other. The coupling strength was accurately
controlled by displacing the common output coupler in and out of
the focal plane of a focusing lens which was inserted in front of
the output coupler~\cite{VarditScience}. Maximum coupling strength
of $5\%$ was achieved when the output coupler was placed at the
focal plane of the lens. The results are presented in
Fig.~\ref{graphs}. Figure ~\ref{graphs}(a) shows the combining
efficiency, which is also a measure of the phase locking between
the lasers, as a function of the coupling strength. As evident,
for the intra-cavity configuration the lasers are strongly phase
locked (high combining efficiency) even at a very low coupling
strength of about 0.5\%, whereas for the outer-cavity
configuration strong phase locking occurs at a higher coupling
strength of 2\%. Figure \ref{graphs}(b) shows the combined output
power as a function of coupling strength. As evident, the output
power is essentially the same over a large range of coupling
strengths.

At the lower coupling strengths, the decrease in output power for
the outer-cavity configuration is due to reduction in phase
locking, whereby more light is lost at the interferometric
combiners. For the intra-cavity configuration, the reduction of
output power is due to less efficient lasing because less light is
reflected from the single common output coupler. These results
extend our earlier results for only two fiber
lasers~\cite{Moti2006, VarditScience} and indicate that it is
possible to upscale phase locking and coherent addition to a
larger number of fiber lasers.

\section{Concluding Remarks}

To conclude, we presented two compact and robust configurations
for efficient coherent addition in free space of four separate
fiber lasers that are arranged in a two-dimensional array, using
two binary interferometric combiners. We obtained high combining
efficiency together with a good beam quality for both
configurations. Although scalability to coherent addition of very
large arrays of lasers is still uncertain~\cite{ManyFibers}, we
believe that our approach could be extended to moderate arrays
size. This could be done by either resorting to more binary
interferometric combiners or two more complex interferometric
combiners where each can coherently add more than two fiber
lasers.

This research was supported in part by the Binational Science
Foundation.

\clearpage

\end{document}